\documentclass[article,pdftex,10pt,a4paper,twocolumn]{article}
\usepackage[numbers]{natbib}
\usepackage{cite}

\usepackage{graphicx}

\begin{document}

%\begin{frontmatter}

% \title[Multipoles in galaxy spin directions]{Multipole alignment in the large-scale distribution of spin direction of spiral galaxies}
\title{Multipole alignment in the large-scale distribution of spin direction of spiral galaxies}

\author{Lior Shamir \\ \small Kansas State University, Manhattan, KS 66502 } % \\ \ead{lshamir@mtu.edu}}
%\author{Lior Shamir}
%\address{Kansas State University \\ 2184 Engineering Hall, 1701D Platt St., Manhattan, KS 66506 \\ Email: lshamir@ksu.edu}

\date{}

\maketitle

\begin{abstract}
Previous observations have suggested non-random distribution of spin directions of galaxies at scales far larger than the size of a supercluster. Here I use $\sim1.7\cdot10^5$ spiral galaxies from SDSS and $3.3\cdot10^4$ spiral galaxies from Pan-STARRS to analyze the distribution of galaxy spin patterns of spiral galaxies as observed from Earth. The analysis shows in both SDSS and Pan-STARRS that the distribution of galaxy spin directions forms a non-random pattern, and can be fitted to a dipole axis in probability much higher than mere chance. These observations agree with previous findings, but are based on more data and two different telescopes. The analysis also shows that the distribution of galaxy spin directions fits a large-scale multipole alignment, with best fit to quadrupole alignment with probability of $\sim6.9\sigma$ to have such distribution by chance. Comparison of two separate datasets from SDSS and Pan-STARRS such that the galaxies in both datasets have similar redshift distribution provides nearly identical quadrupole patterns.
\end{abstract}

% \keywords{Galaxies: general -- galaxies: spiral}

%\end{frontmatter}

% \maketitle
% \ioptwocol

\section{Introduction}
\label{introduction}

Because the spin direction of a spiral galaxy depends on the perspective of the observer, the distribution of the spin directions of spiral galaxies is expected to be randomly distributed in the sky to an Earth-based observer. However, several previous studies showed evidence of non-random patterns of the distribution of spin directions of spiral galaxies \citep{longo2011detection,shamir2012handedness,shamir2013color,hoehn2014characteristics,shamir2016asymmetry,shamir2017colour,shamir2017photometric,shamir2017large,lee2019galaxy,lee2019mysterious,shamir2020asymmetry,shamir2020large}.

First experiments were done with manually annotated galaxies \citep{iye1991catalog,land2008galaxy,longo2011detection}, and showed evidence of asymmetry between the number of galaxies with opposite spin directions. However, manual annotation of galaxies is impractical for analyzing very large datasets of galaxies, and can also be biased by the human perception \citep{land2008galaxy,hayes2017nature}. The ability to automate the galaxy annotation led to far larger datasets, that are less vulnerable to human bias due to the machine-based nature of the annotation. Automatically annotated datasets of galaxy images showed clear and statistically significant asymmetry between clockwise and counterclockwise galaxies in Sloan Digital Sky Survey \citep{shamir2017colour,shamir2017photometric,shamir2020large}. While the first experiments were based on SDSS alone, more recent work using several telescopes showed very good agreement between SDSS and data collected by other telescopes such as Pan-STARRS \citep{shamir2017large}, and Hubble Space Telescope \citep{shamir2020asymmetry}. Other experiments used smaller manually annotated datasets to show patterns of spin directions of galaxies \citep{slosar2009galaxy}, including galaxies that are too far from each other to have gravitational interactions \citep{lee2019mysterious}. Alignment of spin directions has also been shown with quasars \citep{hutsemekers2014alignment}.

Although the observations might be difficult to explain without violating the basic cosmological foundations, explanations to the observations that do not necessarily require the violation of the foundational cosmology have been proposed \citep{yu2020probing,biagetti2020primordial}. Here I use a large dataset of $\sim1.7\cdot10^5$ SDSS galaxies that were annotated automatically by their spin directions to show a possible multipole alignment of the distribution of galaxy spin directions. The patterns identified in SDSS galaxies are also compared to the patterns of galaxy spin directions in a completely separate dataset of Pan-STARRS galaxies.

\section{Data}
\label{data}

Data from two different digital sky surveys were used in this experiment, to allow a comparison between data collected by different telescopes and reduce the probability that the results are driven by an unknown instrumental flaw in a certain telescope system. The first dataset is $\sim1.7\cdot10^5$ spiral galaxies from SDSS annotated by their spin directions, and it is the same dataset used in \citep{shamir2017colour,shamir2017photometric,shamir2017large,shamir2020large}. The galaxies are spiral galaxies taken from the catalog of $\sim3\cdot10^6$ SDSS galaxies classified automatically into spiral and elliptical galaxies \citep{kuminski2016computer}. These galaxies are relatively bright and large, with i magnitude $<18$ and Petrosian radius larger than 5.5''. The  740,908 galaxies identified as spiral galaxies were used, while the galaxies that were identified as elliptical galaxies were removed from the experiment.

In addition to the SDSS dataset, a separate dataset of Pan-STARRS galaxies was used to allow comparisons between the results produced by the two sky surveys. For that purpose, a dataset of 2,394,452 Pan-STARRS galaxies that were labeled by Pan-STARRS photometric pipeline as extended sources in all bands \citep{timmis2017catalog} was used. 

The galaxy images from both datasets were retrieved using the {\it Cutout} service, which both SDSS and Pan-STARRS provide. The images were retrieved in the form of 120$\times$120 JPG images. To ensure that the galaxy fits inside the image, if more than 25\% of the pixels on the edge of the image had grayscale value greater than 125, the image was downscaled by 0.01'' and downloaded again until the number of bright pixels on the edge of the frame was less than 25\% of the total number of edge pixels \citep{kuminski2016computer}. % The initial scale of the image was 0.1'' per pixel, and it was reduced by 0.01'' per pixel in each iteration, until the galaxy fits in the frame as was done in \citep{kuminski2016computer}.

After the images were downloaded, they were separated into galaxies with clockwise spin direction and galaxies with counterclockwise spin direction. That was done by using the Ganalyzer algorithm \citep{shamir2011ganalyzer,ganalyzer_ascl}, as was done in \citep{shamir2012handedness,shamir2013color,hoehn2014characteristics,dojcsak2014quantitative,shamir2016asymmetry,shamir2017colour,shamir2017photometric,shamir2017large}. In summary, Ganalyzer computes the radial intensity plot of each galaxy images, which is a simple transform of the raw galaxy image into an image of dimensionality of 360$\times$35, such that the X axis is the polar angle (in degrees) and the Y axis is the radial distance (in percents of the galaxy radius). The value of the pixel at coordinates $(x,y)$ in the radial intensity plot is the median of the 5$\times$5 pixels around $(O_x+\sin(\theta) \cdot r,O_y-\cos(\theta)\cdot r)$ in the galaxy image, such that $\theta$ is the polar angle and {\it r} is the radial distance. The radial distance is measured in terms of percent of the radius of the galaxy.

\begin{figure}
\includegraphics[scale=0.5]{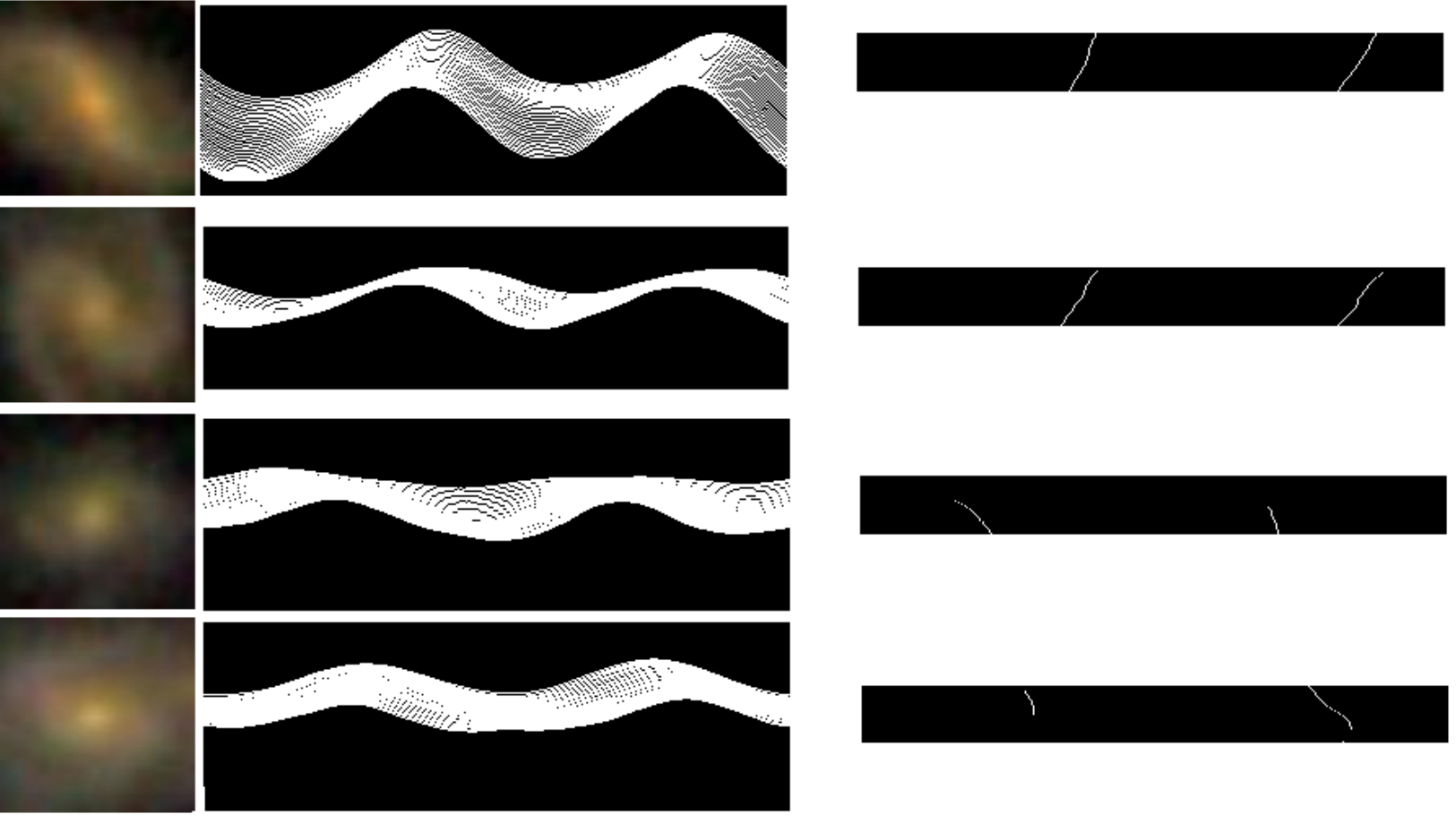}
\caption{Original galaxy images (left), the radial intensity plots (middle), and the peaks detected in the radial intensity plots (right). The sign of the regression of the vertical lines of the peaks determines the curve of the arms of the galaxy, and therefore also the spin direction.}
\label{redial_intensity_plots}
\end{figure}

After the radial intensity plot is computed, peak detection is applied to identify groups of peaks across the horizontal lines of the radial intensity plot \citep{shamir2011ganalyzer}. Since arm pixels are brighter than the sky background, the groups of peaks in the radial intensity plot are the arms of the galaxy. Connecting the peaks detected in different horizontal lines creates vertical lines that correspond to the curve of the arm. Therefore, a linear regression applied to the peaks can reflect the shape of the arm, and the sign of the regression coefficient indicates whether the galaxy has clockwise or counterclockwise spin direction. Figure~\ref{redial_intensity_plots} shows examples of original galaxy images, the transforms into radial intensity plots, and the peaks detected in the radial intensity plots. A through description of the algorithm as well as analysis of its performance can be found in \citep{shamir2011ganalyzer}, as well as in \citep{hoehn2014characteristics,shamir2012handedness}.

% Many spiral galaxies do not necessarily have an identifiable spin direction, and in some cases the algorithm can missclassify a galaxy even if the spin direction is identifiable. To avoid using galaxies that are annotated incorrectly, 

To use galaxies which their spin direction can be determined, only galaxies that had at least 10 identifiable peaks in the radial intensity plot were used, and  galaxies that did not have at least 10 peaks were ignored. Of the peaks that were detected, at least 75\% were expected to be oriented towards the same direction. Galaxies that did not meet these criteria were rejected, and were not used in the analysis. Manual inspection of 200 galaxies with clockwise spin direction and 200 randomly selected galaxies with counterclockwise spin direction showed that 10 galaxies classified as clockwise and 13 galaxies classified as counterclockwise did not have identifiable spin patterns, but none of these galaxies was missclassified.

An important feature of Ganalyzer is that it is a model-driven algorithm that follows defined rules that are driven by the features of the galaxy. As deep neural networks are commonly used for the task of image classification, these methods are based on complex and unintuitive data-driven rules that make it extremely difficult to understand how the classification is performed. Such methods can capture subtle biases in the training set or in the data collection process, and lead to unexplained bias or asymmetry that is difficult to understand due to the complex classification process.

Annotating the dataset of SDSS galaxies by their spin direction using Ganalyzer as described above provided a dataset of 88,273 galaxies with clockwise spin patterns and 86,075 galaxies with counterclockwise patterns. Assuming mere chance probability of 0.5 of a galaxy to be associated with one of the two possible spin directions, the probability to have such separation by chance can be computed using cumulative binomial distribution, such that the number of tests is 174,348 and the success probability is 0.5. The two-tailed probability to have 88,273 or more successes is $P<10^{-7}$. Repeating the experiment after mirroring the galaxies led to identical results, which is expected due to the deterministic and symmetric nature of Galanyzer. Table~\ref{ra_dsitribution_sdss} shows the number of SDSS galaxies in each 30$^o$ redshift range.

% ###add here distribution in different redshift ranges ###

\begin{table}
{
%\footnotesize
\scriptsize
\begin{tabular}{|l|c|c|c|c|}
\hline
RA &     ${cw-ccw}\over{cw+ccw}$   &  \# galaxies  \\ 
\hline
0$^o$-30$^o$           & -0.005    & 27,128   \\
30$^o$-60$^o$        &    0.068 &   18,023  \\
60$^o$-90$^o$        &  -0.031   &   3,477 \\
90$^o$-120$^o$      &    0.011 &   5,302 \\
120$^o$-150$^o$     &   0.022  &   20,820 \\
150$^o$-180$^o$     &   0.005  &   19,585 \\
180$^o$-210$^o$     &  0.012  &    18,466 \\
210$^o$-240$^o$     &  0.021  &    18,894 \\
240$^o$-270$^o$     &  -0.001  &    14,245 \\
270$^o$-300$^o$     &   0.019 &    873 \\
300$^o$-330$^o$     &  -0.04  &    8,351 \\
330$^o$-360$^o$     &   0.016 &    19,184 \\
\hline
\end{tabular}
\caption{The number of galaxies and the asymmetry between clockwise and counterclockwise galaxies in different RA ranges in SDSS.}
\label{ra_dsitribution_sdss}
}
\end{table}

\begin{figure*}
\includegraphics[scale=0.45]{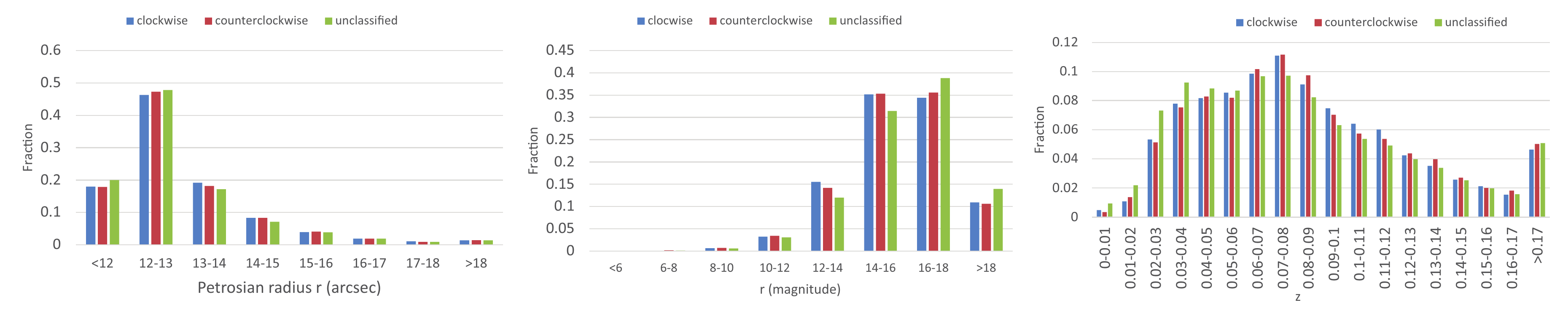}
\caption{Distribution of the exponential r magnitude, Petrosian radius measured in the r band, and the distribution of redshift.}
\label{distribution}
\end{figure*}

Figure~\ref{distribution} shows the distribution of the exponential r magnitude, Petrosian radius measured in the r band, and the redshift of the galaxies that were classified as clockwise, counterclockwise, or could not be classified and were therefore rejected from the analysis. Most SDSS galaxies do not have spectra, and therefore just the subset of 10,281 SDSS DR8 galaxies that had spectra were used for deducing the redshift distribution.

%As the figure shows, while galaxies with higher r magnitude tend to be classified less frequently into clockwise or counterclockwise galaxies, the distribution of the galaxies that could not be classified by {\it Ganalyzer} is largely aligned with the distribution of the galaxies that were classified as clockwise or counterclockwise. Figure~\ref{dataset_accuracy} shows the distribution of galaxies that did not have clear identifiable spin pattern in different redshifts, radii, and r magnitudes.

%\begin{figure}[h]
%\includegraphics[scale=0.6]{dataset_accuracy.pdf}
%\caption{Distribution of the galaxies that did not have clear identifiable spin patterns in different r magnitude, Petrosian radius measured in the r band, and redshift.}
%\label{dataset_accuracy}
%\end{figure}

\begin{figure}[h]
\includegraphics[scale=0.60]{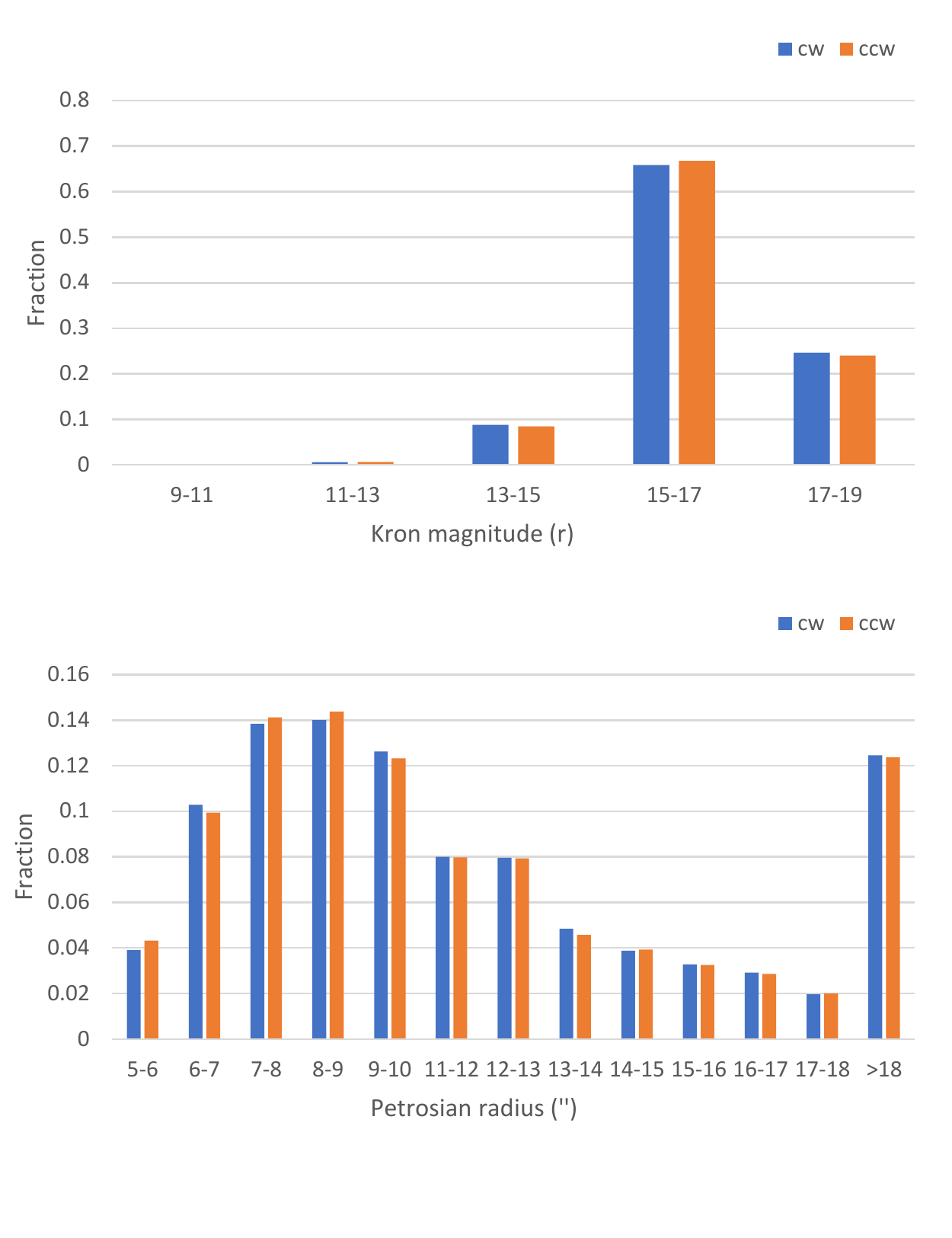}
\caption{The distribution of the Petrosian radius and the r Kron magnitude of the Pan-STARRS galaxies used in the experiment.}
\label{ps_mag_radius}
\end{figure}

\begin{figure}[h]
\includegraphics[scale=0.60]{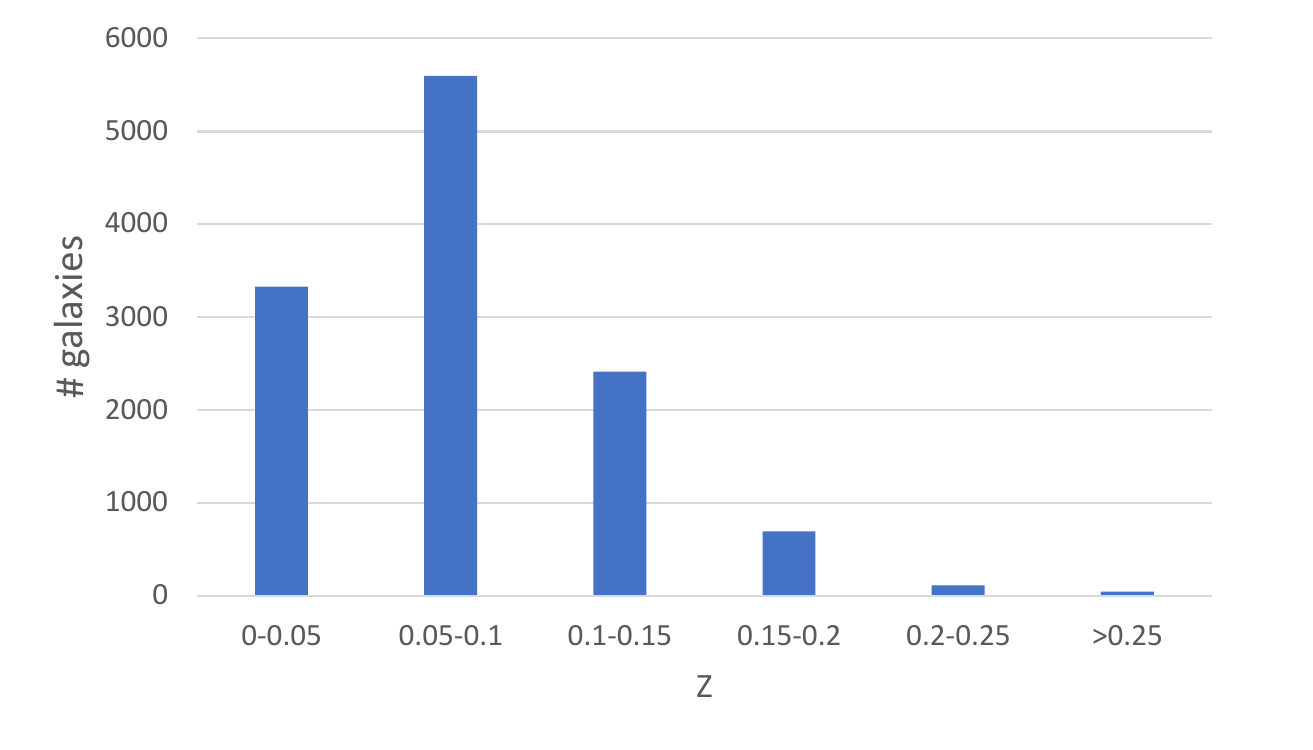}
\caption{The redshift distribution of the Pan-STARRS galaxies. Pan-STARRS is an optical sky survey that does not provide spectroscopy for the objects. Therefore, the redshift distribution was determined by using 12,186 objects that have corresponding spectroscopic objects in SDSS DR8.}
\label{PanSTARRS_z_distribution}
\end{figure}

The application of Ganalyzer to the Pan-STARRS galaxies provided a dataset of 33,028, of which 16,508 had clockwise spin direction and 16,520 had counterclockwise spin. Figure~\ref{ps_mag_radius} shows the distribution of the Petrosian radius measure in the r band and the r Kron magnitude of the galaxies. The redshift distribution of the Pan-STARRS galaxies is shown in Figure~\ref{PanSTARRS_z_distribution}. The redshift distribution is important for making a comparison between the Pan-STARRS and SDSS data, and will be used in Section~\ref{results}. In the case of Pan-STARRS, Ganalyzer was applied to all galaxies that met the criteria described above, and not just to galaxies identified as spiral as was done in SDSS, since no catalog of Pan-STARRS spiral galaxies exists yet. Therefore, the number of Pan-STARRS galaxies with identifiable spin directions is much smaller relative to the size of the initial dataset of Pan-STARRS galaxies.

%  33,028     16508 16520 

The distribution of these galaxies and the difference between clockwise and counterclockwise galaxies in the different RA ranges is shown in Table~\ref{ra_dsitribution_panstarrs}. The table also shows the asymmetry in each 30$^o$ RA range measured by $\frac{cw-ccw}{cw+ccw}$. Consistency between the asymmetries in Tables~\ref{ra_dsitribution_sdss} and~\ref{ra_dsitribution_panstarrs} is not expected, due to differences in the declination and redshift of the galaxies, as will be discussed in Section~\ref{results}.

\begin{table}
{
%\footnotesize
\scriptsize
\begin{tabular}{|l|c|c|c|c|}
\hline
RA &     ${cw-ccw}\over{cw+ccw}$   &  \# galaxies  \\ 
\hline
0$^o$-30$^o$          &   -0.009  &  3,559  \\
30$^o$-60$^o$        &   -0.031  &  2,676  \\
60$^o$-90$^o$        &   -0.015  &   1,698 \\
90$^o$-120$^o$      &   0.041   &   1,099 \\
120$^o$-150$^o$     &  0.025   &   3,473 \\
150$^o$-180$^o$     &  -0.022  &   5,064 \\
180$^o$-210$^o$     &  -0.001   &   5,195 \\
210$^o$-240$^o$     &   0.023  &   4,088 \\
240$^o$-270$^o$     &   0.013  &   1,874 \\
270$^o$-300$^o$     &   0.082  &   429 \\
300$^o$-330$^o$     &   -0.024 &   1,074 \\
330$^o$-360$^o$     &   -0.007 &   2,799 \\
\hline
\end{tabular}
\caption{The number of galaxies and the asymmetry between clockwise and counterclockwise galaxies in different RA ranges in Pan-STARRS.}
\label{ra_dsitribution_panstarrs}
}
\end{table}

\section{Results}
\label{results}

To test for the existence of an axis of asymmetry in the distribution of galaxy spin directions, for each possible $(\alpha,\delta)$ combination the $\cos(\phi)$ of the galaxies were fitted using $\chi^2$ to $d\cdot|\cos(\phi)|$, such that $\phi$ is the angular distance between $(\alpha,\delta)$ and the galaxy, and $d$ is the spin direction of the galaxy (1 clockwise and -1 counterclockwise). To deduce the statistical significance of the axis, each galaxy was assigned a random number within \{-1,1\}, and the $\chi^2$ of fitting the $\cos(\phi)$ of the galaxies to $d\cdot|\cos(\phi)|$ was computed such that $d$ was the randomly assigned spin direction (1 or -1). The $\chi^2$ was computed 1000 times for each $(\alpha,\delta)$. The mean and standard deviation of the 1000 runs were computed for each $(\alpha,\delta)$ combination. Then, the mean $\chi^2$ computed with the random spin patterns was compared to the $\chi^2$ computed with the actual spin directions of the galaxies. The $\sigma$ difference between the mean $\chi^2$ of the random spin directions and the $\chi^2$ of the actual spin directions is the statistical probability of a dipole axis at $(\alpha,\delta)$.

\begin{figure}[ht]
\includegraphics[scale=0.5]{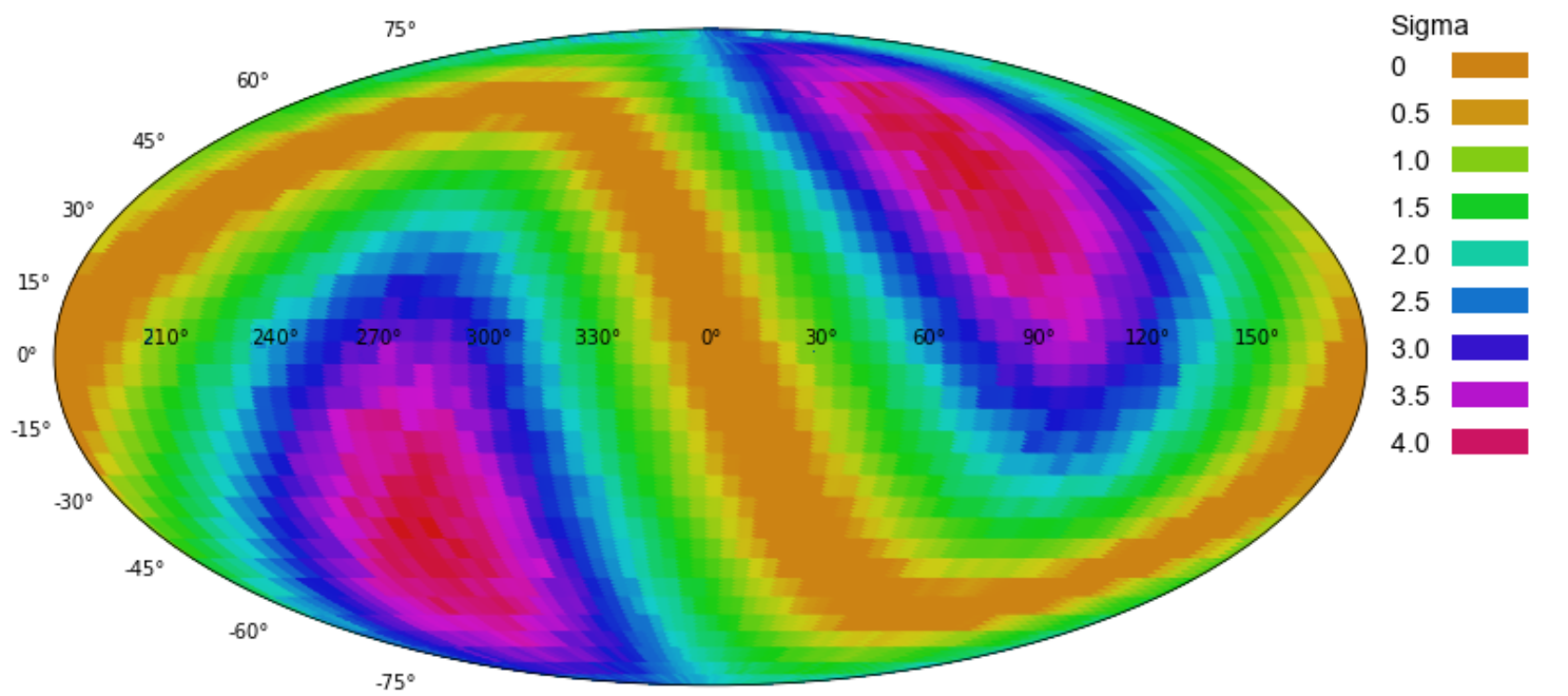}
\caption{The $\sigma$ of possible dipole axes in all combinations of $(\alpha,\delta)$ coordinates using the spin directions of SDSS galaxies.}
\label{dipole_sdss}
\end{figure}

Figure~\ref{dipole_sdss} shows the $\sigma$ of the axis of spin direction asymmetry of all possible $(\alpha,\delta)$ combinations \citep{shamir2020large}. The most likely axis was identified at $(\alpha=88^o,\delta=36^o)$, with $\sigma$ of $\sim$4.34 $(P<0.000014)$. % The probability of an axis to have a 4.34 $\sigma$ from an axis made by randomly assigned spin patterns means that the probability of such axis to occur by chance is $(P<0.000014)$. 
The 1$\sigma$ error of the right ascension of the axis is $(62^o,124^o)$, and the error range of the declination is $(7^o,69^o)$. 

\begin{figure}[ht]
\includegraphics[scale=0.5]{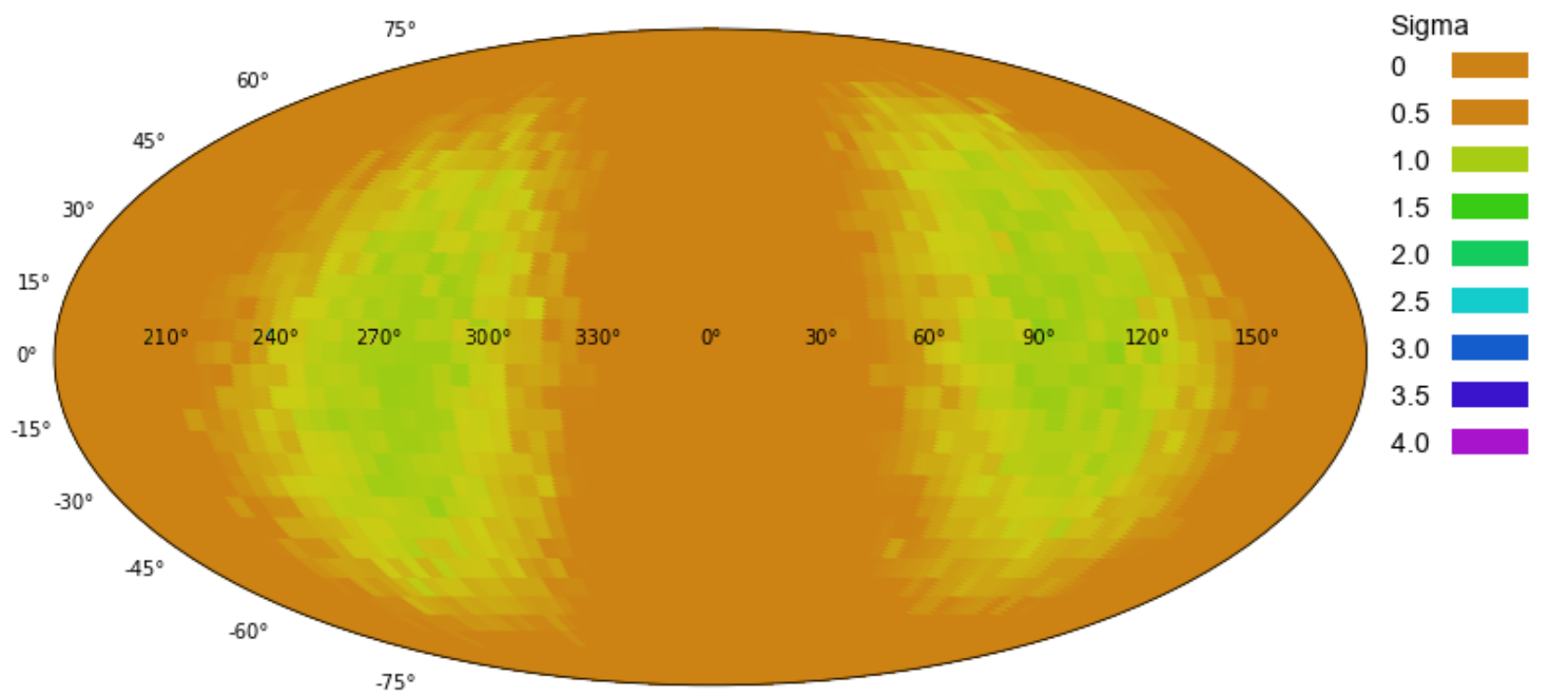}
\caption{The $\sigma$ of the dipole axes in all combination of $(\alpha,\delta)$ coordinates when the SDSS galaxies were assigned with random spin directions.}
\label{dipole_sdss_random}
\end{figure}

Figure~\ref{dipole_sdss_random} shows the probability of an axis when the spin directions of the galaxies are determined randomly. The random spin directions shows no statistically significant pattern, and the maximum probability of the axis is 1.04$\sigma$. The much lower statistical significance indicates that the asymmetry profile is not driven by certain statistical noise.

\begin{figure}[h]
\includegraphics[scale=0.5]{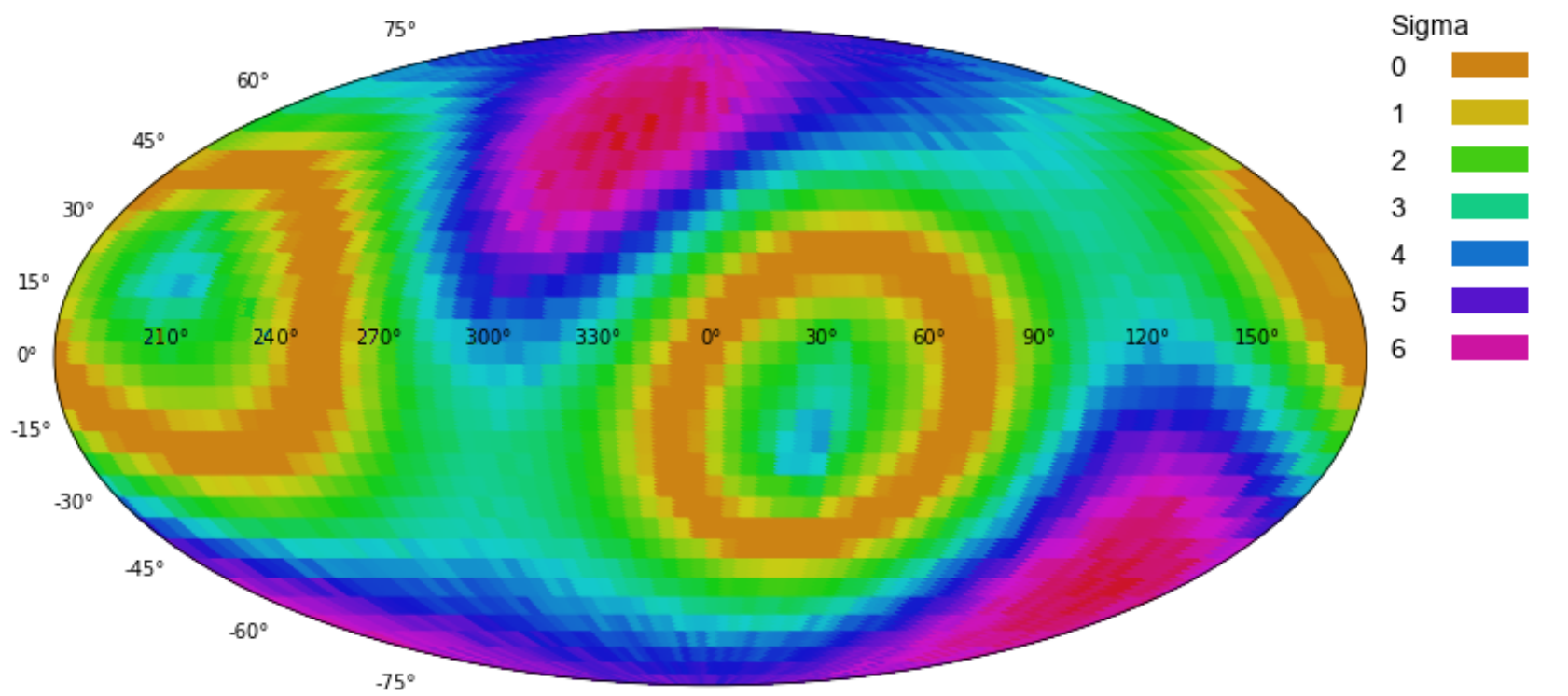}
\caption{The probability of a quadrupole axis in different $(\alpha,\delta)$ combinations.}
\label{quad_sdss}
\end{figure}

\begin{figure}[h]
\includegraphics[scale=0.5]{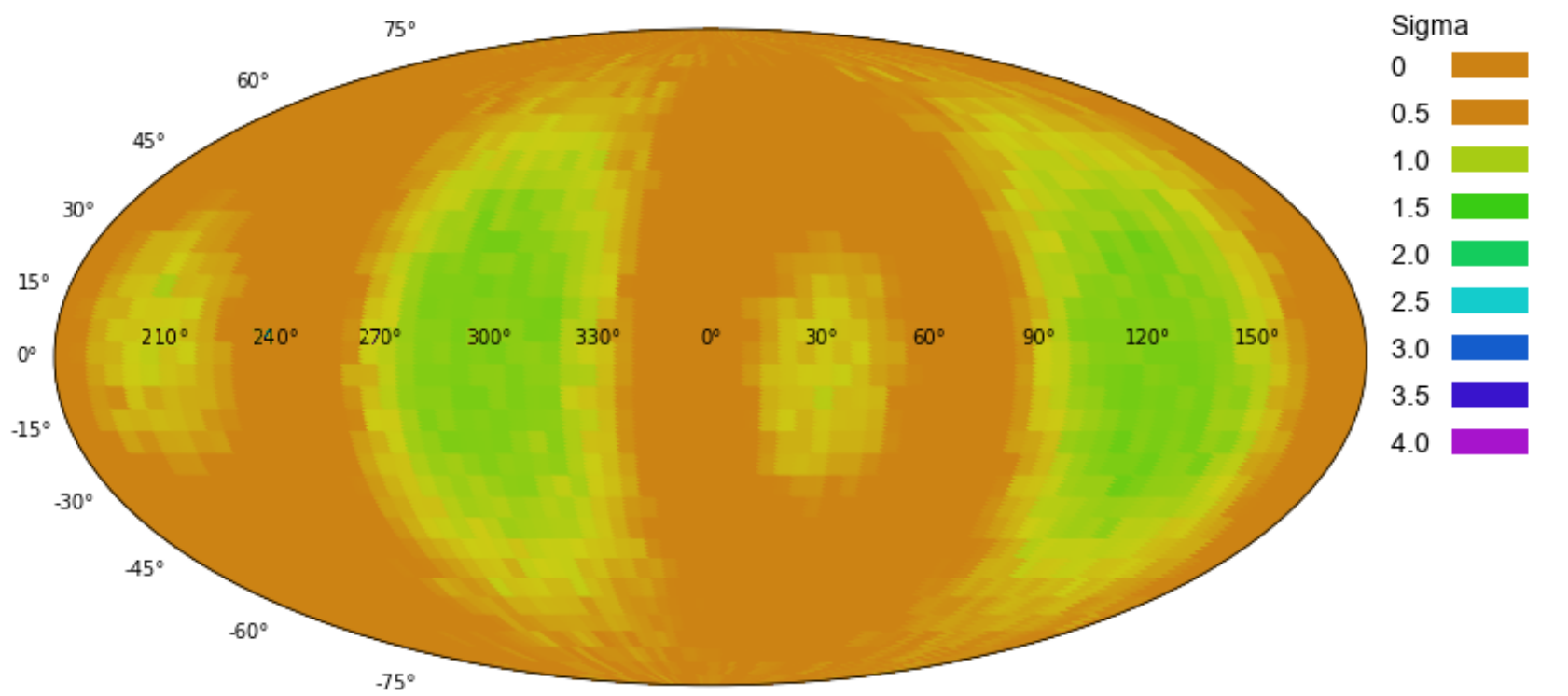}
\caption{The probability of a quadrupole axis in different $(\alpha,\delta)$ combinations when the galaxies are assigned with random spin directions.}
\label{quad_sdss_random}
\end{figure}

Previous observations in the context of CMB anisotropy as observed by COBE, WMAP, and Planck showed evidence of quadrupole alignment \citep{cline2003does,gordon2004low,moss2011induced,zhe2015quadrupole}. Figure~\ref{quad_sdss} shows the $\chi^2$ of fitting the galaxies to cosine $2\phi$ dependence in each possible $(\alpha,\delta)$ combinations. The axis with the highest probability of 6.88$\sigma$ was identified at $(\alpha=333^o,\delta=56^o)$. The 1$\sigma$ error range of the RA is $(292^o,7^o)$, and on the declination the error range is $(36^o,81^o)$. The other axis peaks at $(212^o,17^o)$. Repeating the same experiment when the galaxies are assigned with random spin directions is shown in Figure~\ref{quad_sdss_random}, exhibiting no statistically significant pattern with maximum probability of $1.18\sigma$.

\begin{figure}[h]
\includegraphics[scale=0.5]{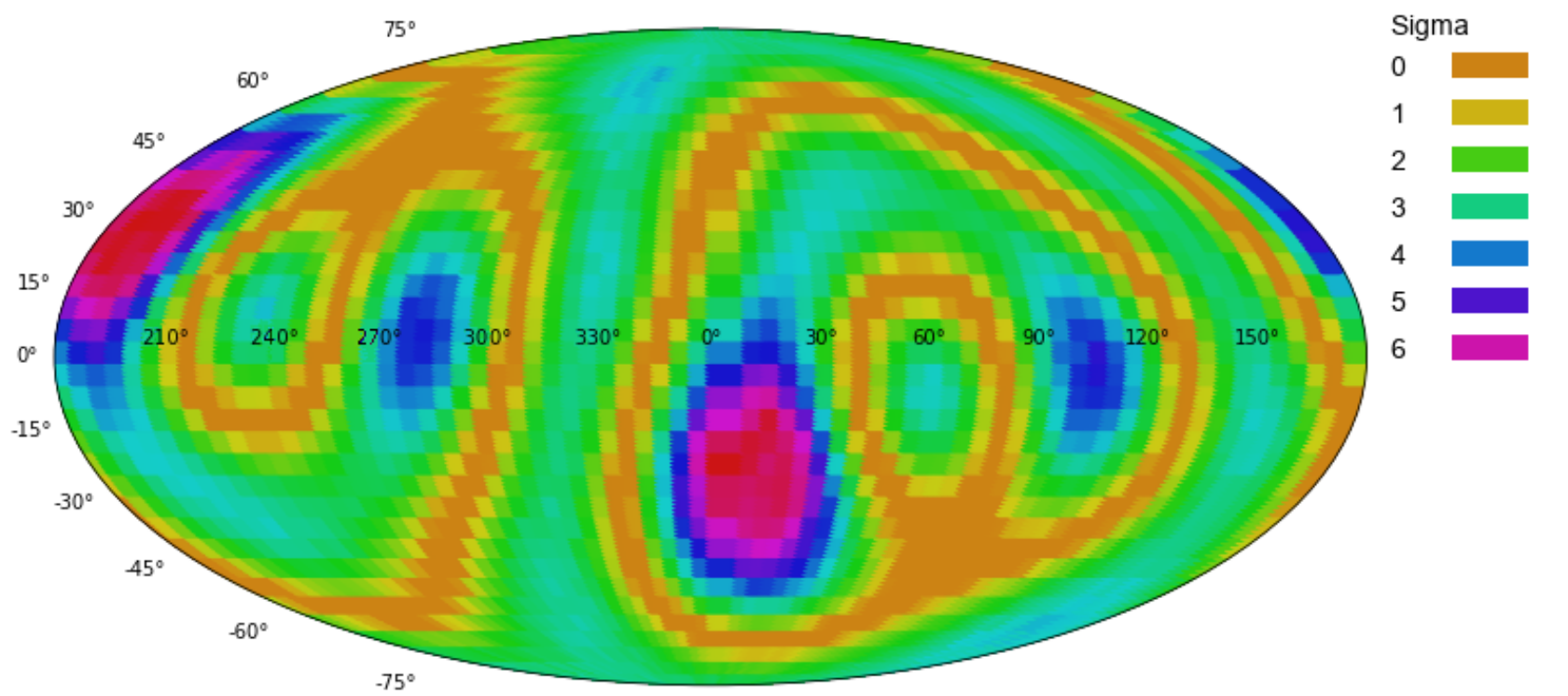}
\caption{The probability of an octopole axis in different $(\alpha,\delta)$ combinations.}
\label{octo_sdss}
\end{figure}

An attempt to fit the galaxy spin directions to octopole alignment is displayed in Figure~\ref{octo_sdss}. The most likely axis is identified at $(\alpha=11^o,\delta=-22^o)$ with probability of $6.41\sigma$. 

\begin{figure}[h]
\includegraphics[scale=0.5]{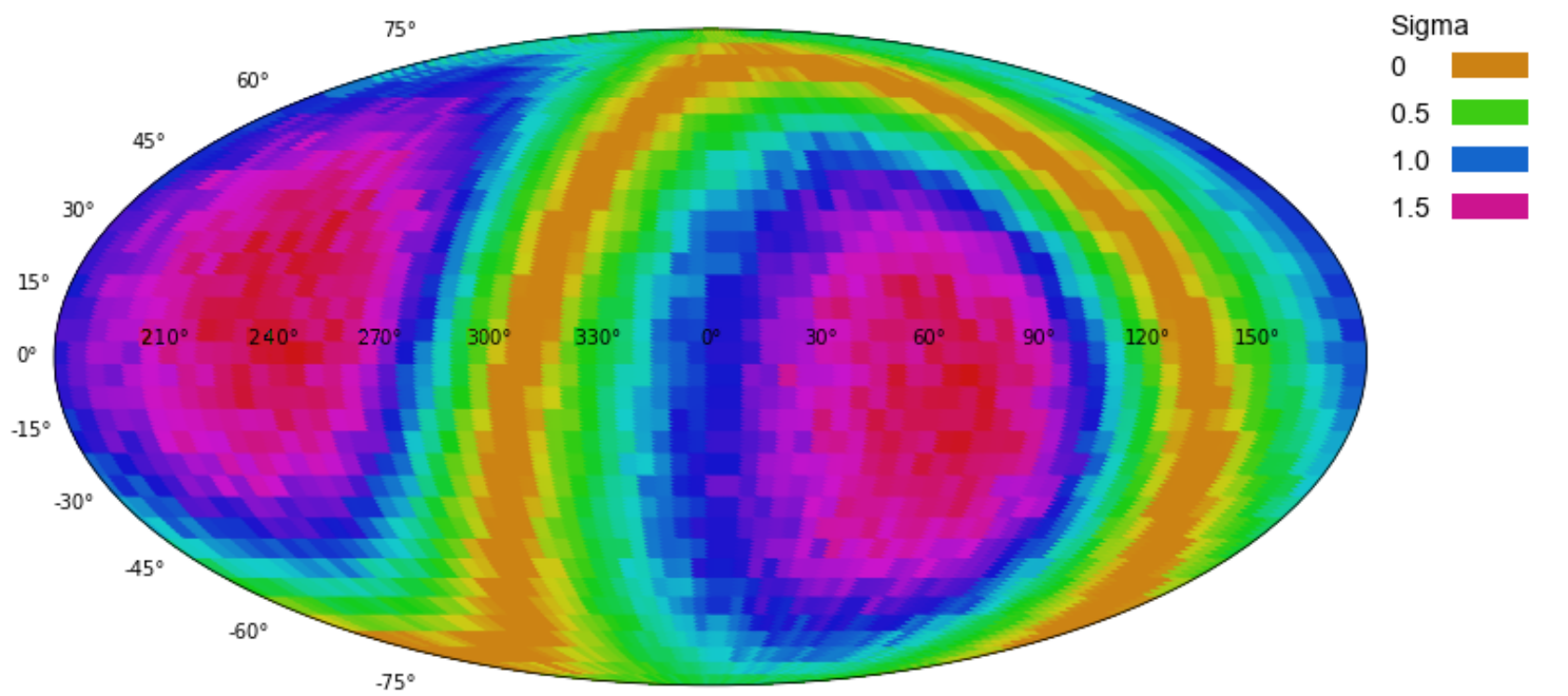}
\caption{The $\sigma$ of the probability of a dipole axis in different $(\alpha,\delta)$ combinations in Pan-STARRS data.}
\label{dipole_ps}
\end{figure}

The patterns of spin direction asymmetry identified by SDSS was compared to the asymmetries identified with the dataset of Pan-STARRS galaxies described in Section~\ref{data}. Figure~\ref{dipole_ps} shows the probability (in $\sigma$) of a dipole axis to happen by chance in each possible $(\alpha,\delta)$ combination. The most likely axis was identified at $(49^o, -3^o)$, with $\sigma$=1.86. The 1$\sigma$ error range is $(3^o,112^o)$ on the right ascension and $(-81^o,52^o)$ on the declination. The figure shows some differences between the large-scale patterns of spin direction asymmetry identified in Pan-STARRS and the patterns identified in SDSS as shown by Figure~\ref{dipole_sdss}, although the differences are well within 1$\sigma$ error.

\begin{figure}[h]
\includegraphics[scale=0.5]{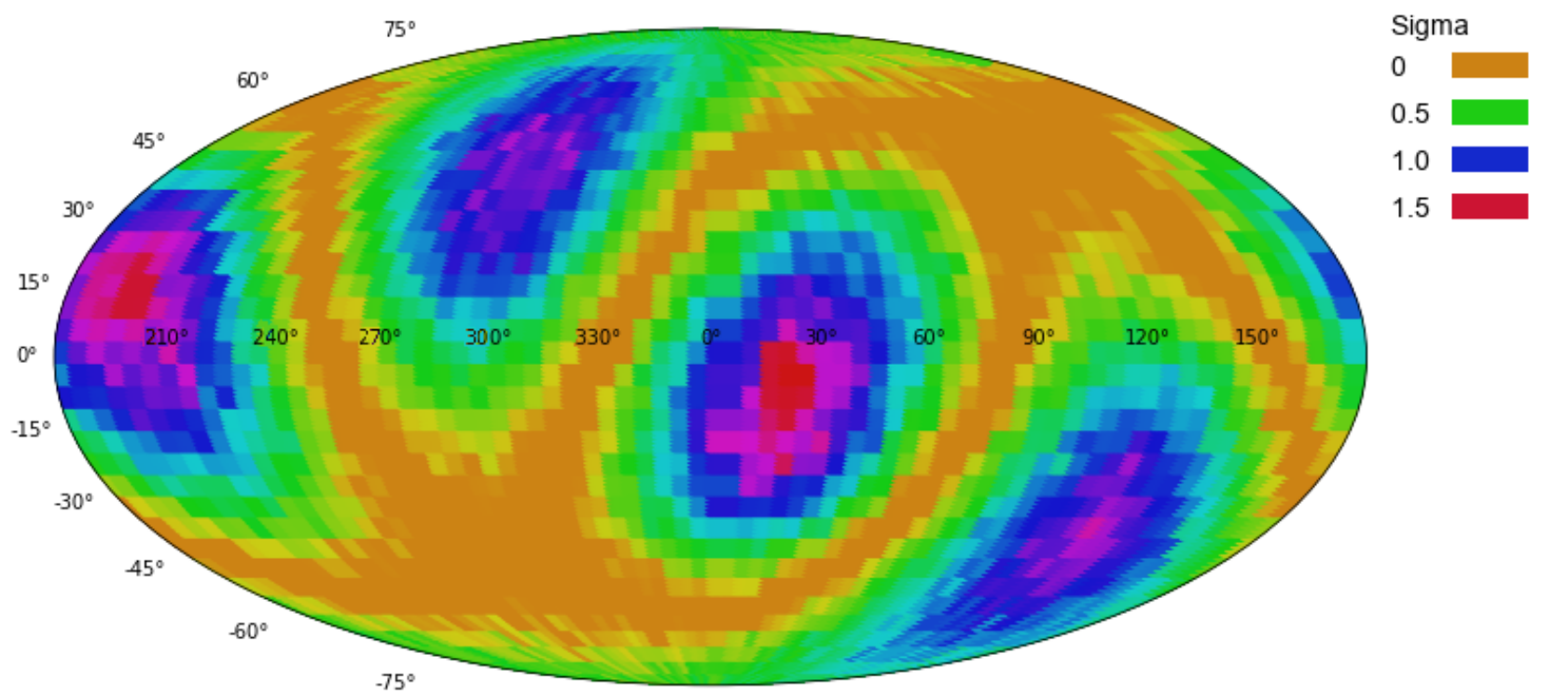}
\caption{The $\sigma$ probability of quadruple alignment to happen by chance in different $(\alpha,\delta)$ combinations in Pan-STARRS data.}
\label{quad_ps}
\end{figure}

The experiments described above with SDSS data showed that the probability of quadrupole alignment of the galaxy spin directions is much higher than the probability of a dipole alignment. Figure~\ref{quad_ps} shows the $\sigma$ probability of quadrupole axis in each possible $(\alpha,\delta)$ combination. The most likely axes are at $(198^o,9^o)$ and $(305^o,40^o)$, with $\sigma$ of 1.65. % and 1.44, respectively. 
These locations are very close to the quadrupole axis identified with the SDSS galaxies.

\begin{figure}[h]
\includegraphics[scale=0.5]{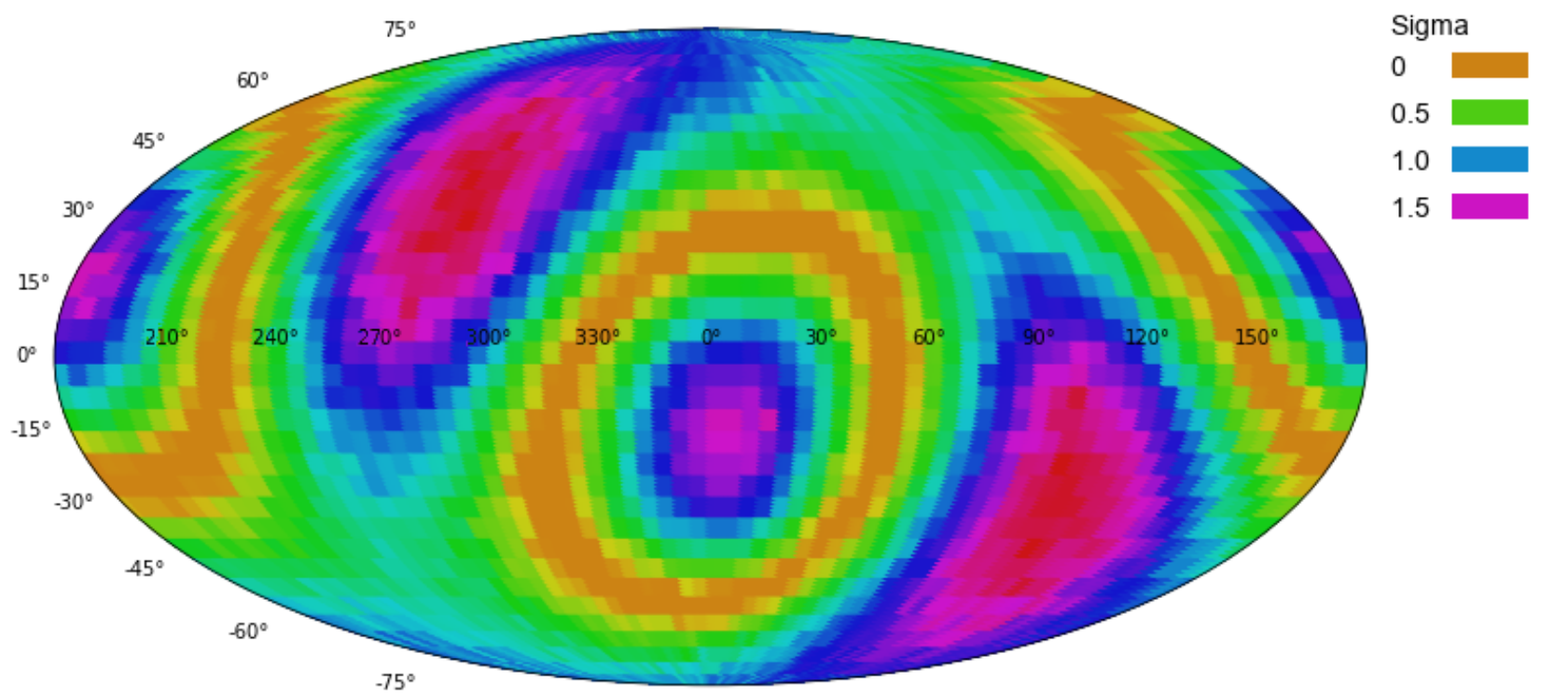}
\caption{The $\sigma$ probability of a quadruple alignment to happen by chance in different $(\alpha,\delta)$ combinations in SDSS galaxies such that the redshift distribution of these galaxies is similar to the redshift distribution of the galaxies in the Pan-STARRS dataset.}
\label{quad_sdss_ps}
\end{figure}

To further compare the large-scale alignment of galaxy spin patterns in SDSS and Pan-STARRS, another dataset of SDSS galaxies was prepared such that the redshift distribution of the galaxies in both datasets was similar. That was done by selecting SDSS galaxies with redshift such that the redshift distribution of the galaxies follows the redshift distribution of Pan-STARRS as shown in Figure~\ref{PanSTARRS_z_distribution}. SDSS DR14 had spectra for 63,693 galaxies, and 38,998 of these galaxies were selected such that the redshift distribution of the galaxies was similar to the redshift distribution of the galaxies in Pan-STARRS as shown in Figure~\ref{PanSTARRS_z_distribution}. Table~\ref{DR14_PanSTARRS} shows the number of galaxies and asymmetry between clockwise and counterclockwise galaxies in each RA range. As the table shows, the asymmetry in the different redshift ranges is more similar to the asymmetry shown in the same RA ranges in Table~\ref{ra_dsitribution_panstarrs}, with Pearson correlation between the asymmetries in the two sky surveys of $\sim$0.55 $(P\simeq0.05)$. The $\sigma$ of quadrupole alignment in the spin directions of these galaxies is shown in Figure~\ref{quad_sdss_ps}. % and Figure~\ref{quad_sdss_z} shows the quadrupole alignment when using all 63,693 SDSS galaxies that had spectra. 

\begin{table}
{
%\footnotesize
\scriptsize
\begin{tabular}{|l|c|c|c|c|}
\hline
RA &   ${cw-ccw}\over{cw+ccw}$ & \# galaxies   \\ 
\hline
0$^o$-30$^o$          &  -0.049 &  2596   \\
30$^o$-60$^o$        &  -0.001 & 1497    \\
60$^o$-90$^o$        &  -0.115 &  61  \\
90$^o$-120$^o$       &  0.024 &  1152  \\
120$^o$-150$^o$     &  0.001 &  8550  \\
150$^o$-180$^o$     &  -0.020 &  10482  \\
180$^o$-210$^o$     &  -0.001 &  7676  \\
210$^o$-240$^o$     &  0.014 &  5892  \\
240$^o$-270$^o$     &  -0.004 &  1018  \\
270$^o$-300$^o$     &  -        &  0        \\
300$^o$-330$^o$     &  -0.081 & 74  \\
330$^o$-360$^o$     &  - &  0  \\
\hline
\end{tabular}
\caption{The number of galaxies asymmetry between clockwise and counterclockwise galaxies in SDSS, such that the galaxies have redshift distribution similar to the redshift distribution of the Pan-STARRS galaxies.}
\label{DR14_PanSTARRS}
}
\end{table}

%\begin{figure}[h]
%\includegraphics[scale=0.3]{quad_sdss_z.pdf}
%\caption{The probability of a quadruple alignment in different $(\alpha,\delta)$ combinations in SDSS galaxies that have spectra.}
%\label{quad_sdss_z}
%\end{figure}

Although the two datasets were collected independently by two different telescopes, both datasets show very similar patterns of quadrupole alignment of the spin directions of the galaxies. The probability of the axis is $\sim1.81\sigma$ and $\sim1.66\sigma$ for the SDSS data and Pan-STARRS data, respectively. Due to the overlap in the footprint of the two telescopes it is expected that some galaxies would be included in both datasets. The number of galaxies included in both the SDSS and the Pan-STARRS datasets is 4,426. Excluding those galaxies such that the two datasets are completely orthogonal has minor impacts on the distribution of the galaxies, and the graph practically does not change when these galaxies are excluded.

\section{Conclusion}
\label{conclusion}

The distribution of galaxy spin directions in SDSS and Pan-STARRS shows patterns in the asymmetry between galaxies with opposite spin directions. Assuming that the asymmetry observed by the two telescopes reflects the real sky rather than a flaw in the telescope systems, the observation can be considered as evidence for large-scale anisotropy. The highest probability was achieved when fitting the distribution of galaxy spin directions to quadrupole, with probability greater than 6$\sigma$. Assigning the galaxies with random spin directions provides no statistically significant patterns.

The method used to annotate the galaxies is a deterministic model-driven algorithm that follows defined rules. It is not based on complex data-driven non-intuitive rules determined during the training process of a deep neural networks, which are very difficult to understand and can capture background noise or subtle biases in the training set. Using randomly assigned spin patterns leads to no statistically significant patterns in the data. Using mirrored images of the galaxies expectedly flips the numbers of clockwise and counterclockwise galaxies, providing also experimental evidence that the algorithm is symmetric. The observation that the asymmetry changes with the direction of observation also shows that a bias in the algorithm is unlikely, since a bias in the annotation algorithm is expected to be consistent in all directions of observation.

The analysis is based on data acquired by the Sloan Digital Sky Survey and Pan-STARRS, and under the assumption that these telescopes and their photometric pipelines are not biased in some unexplained manner that leads to differences between galaxies based on their spin direction. It is difficult, however, to think of a bias that would lead to asymmetry between clockwise and counterclockwise galaxies, as SDSS and Pan-STARRS are not expected to be affected by the spin direction of the galaxy.

The observations reported in this paper agree with previous observations that show asymmetry between galaxies with opposite spin directions   \citep{longo2011detection,shamir2012handedness,shamir2013color,hoehn2014characteristics,shamir2016asymmetry,shamir2017colour,shamir2017photometric,shamir2017large,shamir2020asymmetry,shamir2020large}, or patterns between spin directions of galaxies that are too far to have gravitational interactions \citep{lee2019mysterious}. Cosmological-scale anisotropy was also observed in radio sources \citep{bengaly2018probing} and short gamma ray bursts \citep{meszaros2019oppositeness}, providing evidence of non-uniform distribution that could violate the isotropy assumption of the cosmological principle, although long gamma ray bursts show no statistically significant anisotropy \citep{abreu2012search}. Luminosity-temperature ratio of 313 galaxy clusters also showed violation of the isotropy assumption \citep{migkas2020probing}.

Cosmic Microwave Background (CMB) data also shows evidence of possible cosmological-scale polarization \citep{aghanim2014planck,hu1997cmb,cooray2003cosmic,ben2012parity,eriksen2004asymmetries}, and was fitted to quadrupole alignment \citep{cline2003does,gordon2004low,zhe2015quadrupole}. These observations led to theories that shift from the standard cosmological models \citep{feng2003double,piao2004suppressing,rodrigues2008anisotropic,piao2005possible,jimenez2007cosmology,bohmer2008cmb}. Since the spin patterns of a galaxy as visible from Earth is also an indication of the actual spin direction of the galaxy \citep{iye2019spin}, the large-scale patterns in the distribution of the spin directions can be an indication of a rotating universe \citep{godel1949example,ozsvath1962finite,ozsvath2001approaches,chechin2016rotation}. Other explanations that do not violate the basic cosmological assumptions have also been proposed, such as possible primordial subtle chiral violation exhibited in the current epoch by the asymmetry of the distribution of galaxy spin directions \citep{yu2020probing}, or to parity-breaking gravitational waves \citep{biagetti2020primordial}. As future space-based telescopes such as Euclid and ground-based telescopes such as the Vera Rubin Telescope will allow much more powerful data collection, the asymmetry of the distribution of the spin directions of spiral galaxies can be studied in higher resolution to provide more accurate profiling.

\section*{Acknowledgments}
%This study was supported in part by NSF grants AST-1903823 and IIS-1546079.

Funding for the Sloan Digital Sky Survey IV has been provided by the Alfred P. Sloan Foundation, the U.S. Department of Energy Office of Science, and the Participating Institutions. SDSS-IV acknowledges support and resources from the Center for High-Performance Computing at the University of Utah. The SDSS web site is www.sdss.org.

SDSS-IV is managed by the Astrophysical Research Consortium for the Participating Institutions of the SDSS Collaboration including the Brazilian Participation Group, the Carnegie Institution for Science, Carnegie Mellon University, the Chilean Participation Group, the French Participation Group, Harvard-Smithsonian Center for Astrophysics, Instituto de Astrof\'isica de Canarias, The Johns Hopkins University, Kavli Institute for the Physics and Mathematics of the Universe (IPMU) / University of Tokyo, the Korean Participation Group, Lawrence Berkeley National Laboratory, Leibniz Institut f\"ur Astrophysik Potsdam (AIP), Max-Planck-Institut f\"ur Astronomie (MPIA Heidelberg), Max-Planck-Institut f\"ur Astrophysik (MPA Garching), Max-Planck-Institut f\"ur Extraterrestrische Physik (MPE), National Astronomical Observatories of China, New Mexico State University, New York University, University of Notre Dame, Observat\'ario Nacional / MCTI, The Ohio State University, Pennsylvania State University, Shanghai Astronomical Observatory, United Kingdom Participation Group, Universidad Nacional Aut\'onoma de M\'exico, University of Arizona, University of Colorado Boulder, University of Oxford, University of Portsmouth, University of Utah, University of Virginia, University of Washington, University of Wisconsin, Vanderbilt University, and Yale University.

The Pan-STARRS1 Surveys (PS1) and the PS1 public science archive have been made possible through contributions by the Institute for Astronomy, the University of Hawaii, the Pan-STARRS Project Office, the Max-Planck Society and its participating institutes, the Max Planck Institute for Astronomy, Heidelberg and the Max Planck Institute for Extraterrestrial Physics, Garching, The Johns Hopkins University, Durham University, the University of Edinburgh, the Queen's University Belfast, the Harvard-Smithsonian Center for Astrophysics, the Las Cumbres Observatory Global Telescope Network Incorporated, the National Central University of Taiwan, the Space Telescope Science Institute, the National Aeronautics and Space Administration under Grant No. NNX08AR22G issued through the Planetary Science Division of the NASA Science Mission Directorate, the National Science Foundation Grant No. AST-1238877, the University of Maryland, Eotvos Lorand University (ELTE), the Los Alamos National Laboratory, and the Gordon and Betty Moore Foundation.

\bibliographystyle{apalike}
%\bibliographystyle{Wiley-ASNA}
%\bibliographystyle{elsarticle-num}

%\bibliographystyle{iopart-num}

%\bibliographystyle{phpc}
% \bibliographystyle{aomplain}

%\nocite{*}% Show all bib entries - both cited and uncited; comment this line to view only cited bib entries;
\providecommand{\newblock}{}
\bibliography{asym_axis}

\end{document}